# The Future of Quantum Theory: A Way Out of the Impasse

Ghislain Fourny

*ghislain.fourny@inf.ethz.ch*

*ETH Zürich, Department of Computer Science, Universitätsstrasse 6, 8092 Zürich*

**Abstract**

*In this letter, we point to three widely accepted challenges that the quantum theory, quantum information, and quantum foundations communities are currently facing: indeterminism, the semantics of conditional probabilities, and the spooky action at a distance. We argue that these issues are fundamentally rooted in conflations commonly made between causal dependencies, counterfactual dependencies, and statistical dependencies. We argue that a simple, albeit somewhat uncomfortable shift of viewpoint leads to a way out of the impossibility to extend the theory beyond indeterminism, and towards the possibility that sound extensions of quantum theory, possibly even deterministic yet not super-deterministic, will emerge in the future. The paradigm shift, which we present here, involves a non-trivial relaxation of the commonly accepted mathematical definition of free choice, leading to non-Nashian free choice, more care with the choice of probabilistic notations, and more rigorous use of vocabulary related to causality, counterfactuals, and correlations, which are three concepts of a fundamentally different nature.*

When most scientists think of quantum physics, the first keyword that usually comes to their mind is "randomness". The field is now a century old, and the long and passionate debates initiated in the 20th century have not been settled yet. Disagreements continue to exist, and many different interpretations co-habit without a way to validate them or invalidate them experimentally. Many have come to accept that randomness is inherent to

Nature, and impossibility theorems only reinforce this view. As a consequence, probabilities are central and pervasive in the quantum theory literature, more specifically conditional probabilities where the measurement outcomes (output) are conditioned on the choice of experiments to carry out (inputs). However, it is known that formulas involving probabilities in a quantum setting feel different than classical probabilities, in that they break some rules such as the Law of Total Probability. This leads to a questioning of how conditional probabilities should be interpreted in quantum physics. Finally, the relationship between the measurement outcomes and the choices of measurement axes does not only appear in conditional probabilities, but also in causality considerations. The most widely known discussion is that of the "spooky action at a distance", put forward by Einstein, Podolsky, and Rosen in 1935 [6] to express that the choice of a measurement axis at some first location in spacetime can *affect* in some way the measurement outcome at some remote second location, even though no causation is possible as no information can possibly spread between the two locations. The experimental confirmation of Bell's inequalities [18] only confirmed that the "spooky action at a distance", even if not fully understood, is an integral part of quantum theory and of Nature.

We are now going to go through these three challenges in turn, and sketch what we believe is a conceptually simple way out, even though it requires a change in mindset. Figure 1 summarizes the three challenges and proposed resolution alongside the three kinds of dependencies: causation, counterfactuals and correlations.

First, let us consider indeterminism. Indeterminism means that the result of quantum experiments, i.e., the outcome of a quantum measurement, is undetermined until the measurement actually takes place. Gisin [28] expresses this indeterminism with the law of the excluded middle not applying to the (open) future. This is motivated by several different impossibility theorems. The Kochen-Specker theorem [2], now 50 years old, is known as a major breakthrough in this respect, because it proved the impossibility to completely predict the result of experiments *no matter what the physicist will choose to measure*, and even if the considered experiments are compatible (the observables are said to "commute") with each other.

This insight was best summarized more recently by Conway and Kochen in their Free Will Theorem paper [1]:

> *"If indeed there exist any experimenters with a modicum of free will, then elementary particles must have their own share of this valuable commodity."*

The reader will note that mention of the concept of "free will", also known as "free choice" of a measurement axis by the physicist. Free choice is formally defined by Renner and Colbeck [3] like so: "our criterion for A to be a free choice is satisfied whenever anything correlated to A could potentially have been caused by A." (with "could potentially have been caused", understand here causal structure semantics: "is causally preceded by A"). In other words, and if we use the notations common to the literature on Bell inequalities (the letters a and b for the choices of axes, the letters x and y respectively for the associated outcomes), if we consider a random variable A (the choice) and Λ (typically a wrapper for past initial conditions) located in Minkowski spacetime such that A does not causally precede Λ, and A is *chosen freely*, we have for any values λ and a for Λ and A:

$$P(A=a) = P(A=a | \Lambda=\lambda)$$

To the best of our knowledge, free choice is the last remaining so-called "loop-hole" that has not been closed to date, in the quantum community's efforts to prove that quantum theory cannot be extended to a more complete theory [20]. In fact, free choice is a widely shared belief amongst physicists as well as amongst economists, but we are not aware of any published work that would either formally or experimentally establish it as given. Also, this definition of free choice, which was first documented by Nash [17] as unilateral deviations, is less commonly admitted in other communities such as in neuropsychology or mathematical psychology [9][21]. As a consequence, scientific rigor requires that we should consider the possibility that this assumption might not be realized in Nature, even if it may not be aligned with our intuition.

It is often said that giving up free choice, as for example is the case in less mainstream theories such as Bohmian mechanics [4] -- which Bell was aware of and even inspired by -- then the world is super-determinic and science loses its meaning, as we cannot freely choose what experiments we want to conduct. However, it was suggested by Jean-Pierre Dupuy [5], in a resolution of the Newcomb Paradox [10][21], that there is room in the middle for a weaker, but not trivial definition of free choice:

> *"If there is free will at work in the world, the future could be different from what it will be."*

The involvement of modal logic here is very important, as it refers to possible-worlds semantics, with the formal definition of main modal operators based on a set of possible worlds [12] $\Omega$:

**Necessity** ($\Box A$):
*must* be if it happens in all possible worlds in $\Omega$ ($\Omega \subseteq A$);
**Impossibility** ($\Box \neg A$):
A *cannot* be if it happens in no possible world in $\Omega$ ($\Omega \cap A = \emptyset$);
**Possibility** ($\Diamond A = \neg \Box \neg A$):
A *can/could* be if it happens in at least one possible world in $\Omega$ ($\Omega \cap A \neq \emptyset$);
**Contingency** ($\Diamond A \wedge \Diamond (\neg A)$):
A *may or may not* be if it happens in at least one possible world in $\Omega$ and does not happen in at least one possible world in $\Omega$.

Note that the use of possible worlds in the formalism is completely orthogonal to the philosophical question of whether they are real (e.g., the Many-Worlds interpretation [23]) or not.

Dupuy's definition of free choice, which we call non-Nashian free choice, is thus weaker, as it requires the existence of a possible world in which the choice made by the experimenter is different, but does not require statistical independence, and does not require counterfactual independence either. Also, this definition is formally distinct from super-determinism, as Dupuy in fact argues that it gives a more solid basis for the philosophy of ethics. The two different definitions of free choice are shown visually, in their mindsets, on Figure 3.

Another problematic aspect of the mainstream definition of Nash-like free choice is that it conflates correlations and causality, by deriving statistical independence from causal independence in the context of human choices. While of course the statistical independence of our choices is certainly possible, and is even a good approximation of our reality, experience and intuition, it would be a formal mistake to conclude that it is also *necessary* and that this is the only sound model available.

Our first insight and suggestion is that giving up the definition of free choice of a decision in terms of statistical independence of anything that is not causally preceded by the decision is possible without diving into super-determinism. Approaching quantum theory with the weaker, non-Nashian form of free choice ("I could have acted otherwise") circumvents all known impossibility theorems related to the non-extensibility of quantum physics, as they all assume Nash free choice to the best of our knowledge. This re-opens the door for avenue of research in this direction.

Let us now move on to conditional probabilities. Formally, probabilities are defined based on a measure (the probability function) on a measurable space of elementary events Ω endowed with a so-called σ-algebra $\mathcal{F}$ on its subsets (it basically means that it is closed under countable intersections and countable unions). For two events A and B:

$$P(A|B) = \frac{P(A \wedge B)}{P(B)}$$

Conditional probabilities are very common in the quantum literature, in particular when conditioning the outcome of experiments on the choice of axis, e.g., in Bell-inequality setups [16][18] with two physicists, Alice and Bob, each picking an axis (random variables A, B) and obtaining an outcome (Random variables X for Alice, and Y for Bob). The setup is illustrated on Figure 2.

The probability that Alice chooses a and obtains x, and Bob chooses b and obtains y is then:

$$P(X = x \wedge Y = y | A = a \wedge B = b)$$

which is often abbreviated as the common abuse of notation (acceptable when it is clear from the context which value corresponds to which random variable):

$$P(x, y | a, b)$$

Conditional probabilities work quite well in the Bell inequalities literature. However, the problems with conditional probabilities start arising in the quantum literature involving Kochen-Specker-contextuality [2]. Put simply, the issue is that it is generally not possible to find a joint probability distribution common to all observables [5], i.e., they cannot be consistently "glued" together. From a scientific viewpoint, this is preoccupying, as it means that any formalism involving conditional probabilities in a contextual quantum setting is not firmly based on the sound foundations of a properly defined $\Omega$ set with its σ-algebra of subsets $\mathcal{F}$, and a properly defined probability measure on ($\Omega$, $\mathcal{F}$) -- the joint probability.

The way back to a sound formalism that we suggest here is purely conceptual: these conditional probabilities are, in fact, not conditional probabilities. Rather, they are so-called *probabilities of subjunctive conditionals* that are incorrectly expressed with conditional probability notation, thus leading to confusion. The distinction between the two was first pointed to by Stalnaker [15] in a letter to Lewis, as far as we are aware of. Probabilities of subjunctive conditionals express counterfactual dependencies and are not restricted by the same rules as conditional probabilities and thus do not preclude the existence of a joint probability space.

In order to understand this, we must go back to modal logic and possible worlds, and in particular counterfactuals. Counterfactuals first received formal attention in the 1960s and

1970s by David Lewis [11] and Robert Stalnaker [14]. There are several possible definitions, and let us focus on one of them for the purpose of the letter at hand. Given a specific world, i.e., a specific element ω of Ω, the counterfactual binary operator ⟩, based on two hypothetical events A and B (in the sense that they are elements of the σ-algebra $\mathcal{F}$ and not necessarily endowed with a spacetime location), is defined like so.

$$A \rangle B$$

is true in ω if, in the closest world to ω in which A is true (ω ∈ A), B is true as well (ω ∈ B). This has two implications: first, it presupposes that Ω is endowed with a distance structure, i.e., is a metric space. Second, ⟩ is not a truth functional (like ∧ and ¬ are), in that it is not only dependent on the truth value of A and B in the actual world considered (here ω), but also depends on the counterfactual structure derived from the metric (note that the same applies to the necessity and possibility operators □ and ◇). Note that A might be false in ω, which is why such statements are called counterfactuals (contrary to the facts). In the special case that A is true in ω, A ⟩ B holds if and only if B holds. Counterfactual statements do not know the arrow of time and counterfactual dependencies can occur in any direction: there is no causation here.

Counterfactuals are omnipresent in quantum theory: semantically, they appear everytime a subjunctive conditional tense is used, for example "*If* Alice *had measured* the spin along the Y axis instead of the X axis (for which she found -1), then the measured spin *would have been* 1". All these statements fundamentally involve a counterfactual binary operator on the possible worlds: Alice measured along the X axis and obtained a spin of -1; In the closest possible world in which she measured along the Y axis, she obtained a spin of 1.

In particular, we suggest concretely that a conditional probability expressed as:

$$P(X = x, Y = y | A = a, B = b)$$

ought to be expressed formally as the probability of a subjunctive conditional (note the left-right swap):

$$P(A = a \wedge B = b \,\rangle\, X = x \wedge Y = y)$$

meaning that, instead of saying that we compute the probability that the measurement outcomes are x and y conditioned on the chosen measurement axes being a and b, what we are actually computing (and the actual, intended meaning of the quantum framework) is the probability that we are in a world such that, in the closest world from it in which the chosen measurement axes are a and b, the observed outcomes are x and y. The parameter passed in P in the above probability of a subjunctive conditional is itself an event that may or may not hold in any world (the counterfactual implication operator maps two events to another event, given a specific world); as a consequence, the definition is sound, and based on a well-defined properly defined $\Omega$ set with its σ-algebra of subsets $\mathcal{F}$, a properly defined probability measure on $(\Omega, \mathcal{F})$ -- the joint probability.

Some "laws" applying to conditional probabilities such as total law of probability *need not generally apply* to probabilities of subjunctive conditionals, for example:

$$P(A = a) \neq \sum_n P(B = b_n \,\rangle\, A = a) P(B = b_n)$$

in general.

The way that probabilities of subjunctive conditionals behave [13], in turn, is compatible with the counterintuitive behavior of quantum systems precisely because it is not bound by the classical rules of conditional probabilities.

Another problematic aspect of the use of conditional probabilities is that it conflates correlations and counterfactual dependencies, by modeling a formal reasoning on counterfactuals (alternative choices of measurement axis) with statistical correlations. Such a model is an approximation of reality that works at macroscopic scales, but loses its validity at microscopic scales, as the proofs that Nature is contextual force us to notice.

As a consequence, our second insight and suggestion is to abandon the conditional probability notation in favor of a probability of subjunctive conditionals, in order to avoid confusion and, above all, to retrieve a sound framework with a well-defined probability space and joint probability.

Let us now move on to the spooky action at a distance. This dates back to the 1935 paper by Einstein, Podolsky and Rosen [6], when the question was asked whether the quantum-mechanical description of reality is a complete description of reality. It is widely believed today that Einstein was incorrect in believing that a more complete description exists. The spirit of this letter is to support Einstein's view that there exists such a more complete description, with material consequences on our prediction capabilities. A look at this paper is very informative, and also shows the high degree of mastery of the authors:

> *"If, instead of this, we had chosen another quantity, say B, having the eigenvalues $b_1$, $b_2$, $b_3$, and eigenfunctions $v_1(x_1)$, $v_2(x_2)$, $v_3(x_3)$, we should have obtained, instead of Eq. (7), the expansion*
>
> $$\phi(x_1, x_2) = \sum_{s=1}^{\infty} \varphi_s(x_2) v_s(x_1)$$
>
> *where $\phi_s$'s are the new coefficients."*

One can only notice the cautious use of the subjunctive conditional tense, which implies a counterfactual statement as described above, and not a causal relationship. It is unfortunately widespread in the quantum literature to use causal terminology for what are in fact counterfactuals. Examples of inappropriate use of causal terminology is *"change"*, *"effect"*, *"influence"* when counterfactual dependences are meant, i.e., when the two sides are *not in the same possible world*. Formally, an *effect* requires that an event causally precedes another event.

Likewise, a *change* of the choice of measurement axis requires that some choice is made at a time $t_1$ (and the measurement carried out), and then another choice is later made at a time $t_2>t_1$ (and another, different measurement is carried out, on the same system, or on another copy of the system in the sense that it was prepared with the same protocol). This is not the same as saying that what would have happened if one *had (hypothetically) chosen a different axis*. Likewise, computations of correlations are correct if considering, say, millions of measurements that *were actually carried out and duly logged*, which is not the same as considering measurements that *could have been, but in fact were not, carried out*.

Having this in mind, it follows that the spooky action at a distance is not an action: there is no cause and effect, there is no influence, but there is only a counterfactual dependence: *if Alice had chosen a different measurement axis, then Bob would have measured this outcome instead.* This in turns formally means that in the closest world to us in which Alice picks a different measurement axis, Bob measures this other outcome. It thus follows as a corollary that this dependence is not spooky and is rather a statement on the counterfactual structure implied by the rules of quantum physics.

Another problematic aspect of the improper use of causal terminology for counterfactual dependencies is that it conflates causation with counterfactual dependencies. *Conflation* and not *confusion*, because defining causality in terms of counterfactuals is, in fact, done on purpose and by design (A causes B when, if A were not true, B would not be true either). This definition has a pragmatic origin and is actually useful, because it is instrumental to data science for decision making [7][8][27] and this alternate definition also leads to the concept of retrocausality [26] or backtracking counterfactuals [11]. However it is to be distinguished from causation in the relativistic sense.

It is informative to understand that this intended conflation between causation and counterfactuals is a direct consequence of the axiom of Nash-like free choice as the statistical independence of human decision from anything that they could not have caused. This is because, under this mainstream view of free choice, in the closest possible world in which the decision is different, the entire past light cone is exactly the same as in the actual world. For this reason, it is tempting to say that one *changes the choice of measurement axis* while keeping the entire past light cone fixed. Weakening free choice to a non-Nashian version as explained further up involves accepting that such a possible world, in which the present is different but the past is exactly the same, *might not exist in the first place*, which is also a way of restoring a deterministic approach -- in the sense that the past determines the present and future -- but without super-determinism -- in the sense that the past, present and future *could have been different*.

Our third insight and suggestion is to carefully distinguish between causation and counterfactual dependence, ideally by no longer describing counterfactual dependencies with causal terminology.

We hope that the considerations made in this letter stimulate further research in quantum theory and convince more scientists to not give up on Einstein's belief that quantum theory is, in fact, extensible into a sound, wider theory with more predictive power. These considerations involve a shift of viewpoint that, in no way, goes against any previous research. Rather, we are advocating that reformulating previous and future quantum theory research insights rigorously by carefully distinguishing between causation, counterfactual dependencies and correlations is likely to act as a catalyst for further advancing the field of quantum foundations, because it forces us to think in a more precise fashion that is conducive to progress. The quantum world might be simpler than we think, but this is something we will only be able to notice if we think clearly.

## Caption of Figure 1

We see the current challenge of quantum theory as arising from a three-way confusion between causation, counterfactuals and correlations: the indeterminism issue is caused by a conflation of causal independence and statistical independence leading to the Nash-like free choice assumption. The issue with probability semantics arises from a confusion of correlations (conditional probabilities) and counterfactuals (probabilities of subjunctive conditionals). The issue with the spooky action at a distance arises from a conflation of causation and counterfactuals and improper use of causation terminology (e.g., "change", "influence") for counterfactual dependencies.

The way out lies in a careful distinction between three kinds of dependencies and a rigorous use of terminology related to them.

## Caption of Figure 2

The experiment behind the Bell inequality, as typically shown in the literature. This specific layout is inspired by a talk by Matthew Pusey in the ETH workshop "Time in Quantum Theory". Two maximally entangled particles are distributed to Alice and Bob who then take them to their labs. They each choose a measurement axis (for spins, they are known as X, Y or Z -- the Pauli matrices), modelled with random variables A (with value a) and B (with value b). The measurement outcomes are modelled with random variables X (with value x) for Alice and Y (with value y) for Bob. A and Y as well as B and X are spacelike-separated, and this is where the spooky action at a distance occurs. Commonly, free choice for A and B is commonly assumed, meaning that they are statically independent from a hypothetical past commonly denoted as Λ. It is this assumption that forces an indeterministic view of the measurement outcomes.

## Caption of Figure 3

The commonly accepted Nash-like free choice assumption in the physics community can be expressed in terms of the existence of a possible world with exactly the same past light cone, but a different decision regarding the choice of measurement -- to the best of our knowledge, this is also the assumption behind the growth of causal sets [24][25]. We suggest to follow Dupuy's approach to relax this assumption down to allowing another world with a different decision (weaker free choice), but accepting that the past *would have been different* (note the tense, i.e., this is not causation but a counterfactual statement).

**Figure 1**

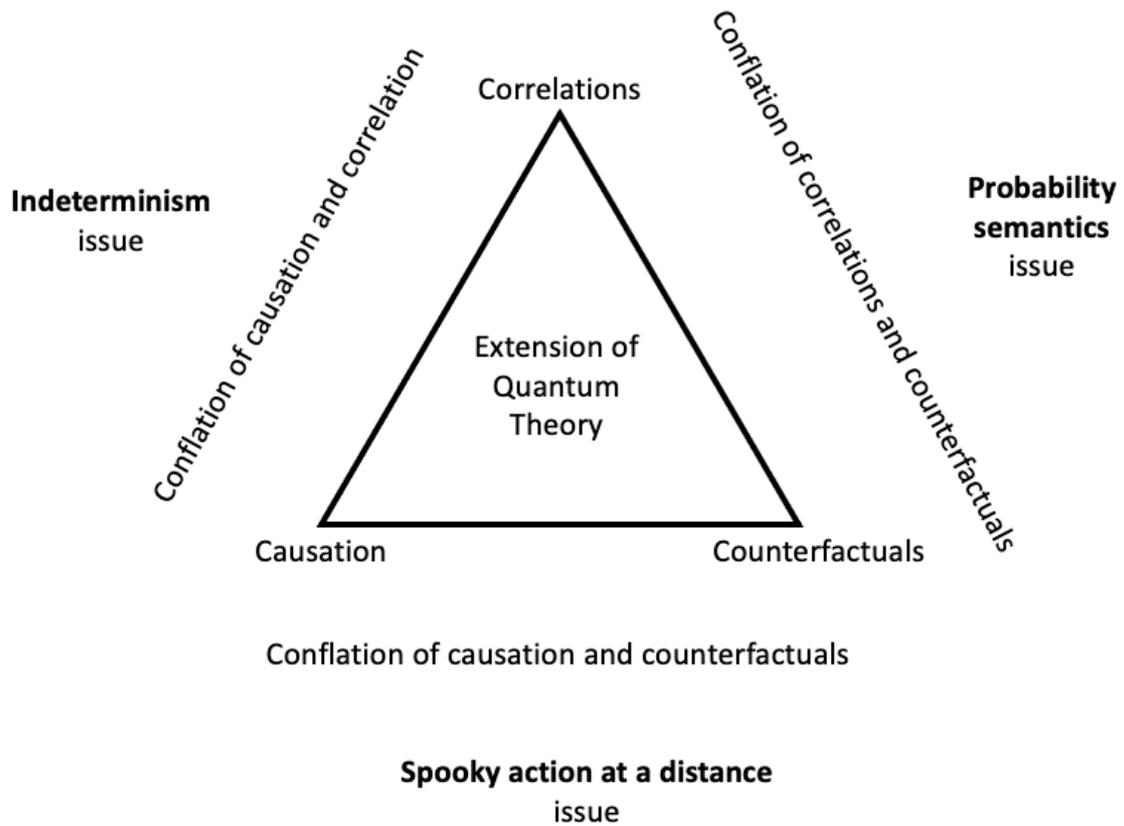

**Figure 2**

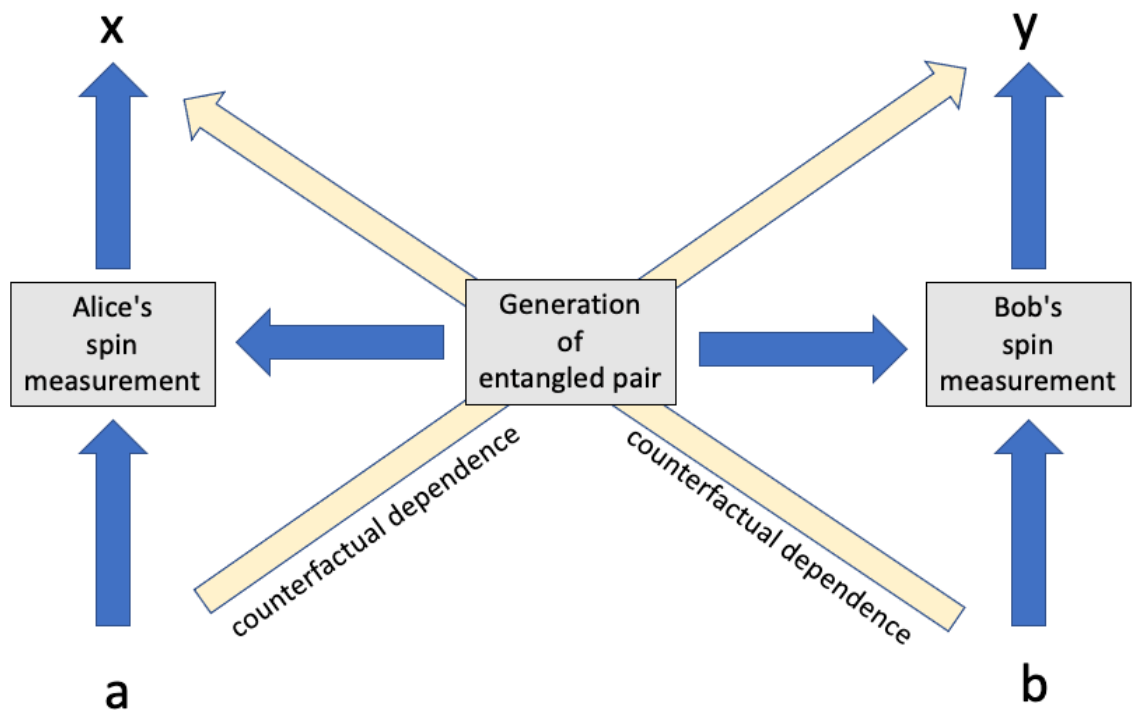

## Figure 3

**Nash** form of free choice

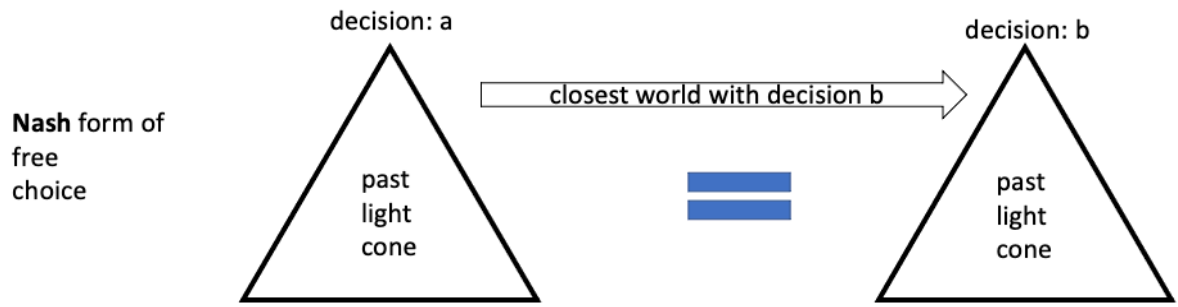

**Non-Nashian** form of free choice

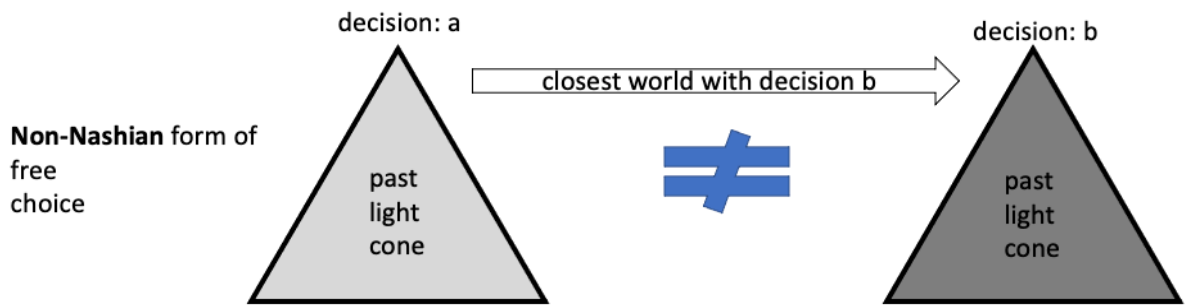